\documentclass[11pt]{article}

\usepackage[utf8]{inputenc}
\usepackage[T1]{fontenc}
\usepackage{lmodern}
\usepackage[english]{babel}
\usepackage{amsmath,amssymb,amsthm}
\usepackage{graphicx}
\usepackage{epsfig}
\usepackage{epstopdf}
\usepackage{subfig}
\usepackage{float}
\usepackage{booktabs}
\usepackage{hyperref}
\usepackage[margin=2.5cm]{geometry}
\usepackage{xcolor}
\usepackage{soul}
\usepackage{changepage}

\newcommand{\ket}[1]{\left|#1\right\rangle}

\renewcommand{\hl}[1]{#1}

\providecommand{\highlighting}[1]{#1}

\newcommand{\address}[1]{\date{}}

\newcommand{\reftitle}[1]{}
\newcommand{\PublishersNote}[1]{}
\newcommand{\authorcontributions}[1]{\section*{Author Contributions}#1}
\newcommand{\funding}[1]{\section*{Funding}#1}
\newcommand{\dataavailability}[1]{\section*{Data Availability Statement}#1}
\newcommand{\acknowledgments}[1]{\section*{Acknowledgments}#1}
\newcommand{\conflictsofinterest}[1]{\section*{Conflicts of Interest}#1}

\newcommand{\firstpage}[1]{}
\newcommand{\pubvolume}[1]{}
\newcommand{\issuenum}[1]{}
\newcommand{\articlenumber}[1]{}
\newcommand{\pubyear}[1]{}
\newcommand{\copyrightyear}[1]{}
\newcommand{\externaleditor}[1]{}
\newcommand{\datereceived}[1]{}
\newcommand{\daterevised}[1]{}
\newcommand{\dateaccepted}[1]{}
\newcommand{\datepublished}[1]{}

\renewenvironment{abstract}{%
  \begin{quote}\noindent\textbf{Abstract:}\ignorespaces
}{%
  \end{quote}%
}

\newlength{\extralength}
\title{Complexity of Nuclear States for $^{48}$Ca}

\author{\parbox{\textwidth}{\centering
L. L\'opez-Hern\'andez$^{1}$, D. A. Lara Bustillos$^{1}$, 
Carlos E. Vargas$^{1,2,*}$ and V. Vel\'azquez$^{1}$\\[4pt]
$^{1}$ Facultad de Ciencias, Universidad Nacional Aut\'onoma de M\'exico, 
Apartado Postal 70-543, Mexico City 04510, Mexico\\
$^{2}$ Facultad de F\'isica, Universidad Veracruzana, Paseo No. 112, 
Desarrollo Hab. Nvo. Xalapa, Xalapa 91097, Veracruz, Mexico\\[4pt]
$^{*}$ Correspondence: cavargas@uv.mx
} }
\begin{document}
\date{}
\maketitle

\begin{abstract}
~In complex systems theory, there are different ways to describe a system in terms of information, such as emergence (Shannon entropy), self-organization, and complexity. These measures provide information about the dynamic behavior of a complex system. We study the differences in entropy and complexity for many-body systems undergoing a transition from a regular to a chaotic regime. To do this, we analyze the eigenvectors of the $^{48}$Ca nucleus for different quadrupole-type two-body interactions. We obtain the eigenvectors by diagonalizing the two-body Hamiltonian for $^{48}$Ca using the Antoine code. We then calculate the entropy and complexity for the different quadrupole-type interactions. The differences found in information entropy and complexity are clear when comparing a regular system with a chaotic one. We find that the complexity of the regular and chaotic states of $^{48}$Ca shows differences associated with its internal interactions.
\end{abstract}


\section{Introduction}

Complex systems are systems composed of many units that interact with each other, giving rise to behaviors that are challenging to model due to intricate interactions among their components and with their environment. These interactions result in a wide range of properties, including nonlinearity, emergence, robustness, and adaptability, among others. The study of complex systems is relevant in various scientific domains, with applications in biology, economics, the social sciences, climatology, medicine, and other fields. In physics, the study of complex systems has evolved from foundational concepts in statistical mechanics and chaos theory, and it continues to expand by exploring how these concepts can be applied to address real-world challenges and uncover the underlying patterns governing diverse systems \cite{Rickles,Benjamin,Hsuan}. {We may start our study of complexity in nuclear systems with the following definition: a system governed by specific laws and subject to prescribed initial and boundary conditions is complex if the interactions within it give rise to new information that was not initially accounted for.} Building on the above, our interest is to study the Shannon entropy and complexity of nuclear states of $^{48}$Ca, which are known to exhibit statistical transitions in their spectra, ranging from $1/f$ noise associated with quantum chaos \cite{Bohigas,Berry,MW,Chirikov} to $1/f^{2}$ noise \cite{Relano, Landa}.

The structure of this article is as follows: in Section two, we will describe the energy spectra of $^{48}$Ca that we will
be working with. Subsequently, in Section three, we will explain how to calculate Shannon entropy and its complexity
 and provide an overview of its behavior depending on the two-body quadrupole interaction. In Section four, we
establish a distance measure to an equiprobable system, namely a system independent of interactions, and compute the complexity of nuclear states. Finally, in the last section, we present our conclusions.

\section{Nuclear States of $^{48}$Ca} 

For the study of Shannon information and complexity, we use the energy levels and eigenvectors of the $^{48}$Ca nucleus. This nucleus has been used in theoretical studies to describe the statistical transitions of its energy levels as a function of the quadrupole interaction \highlighting{\cite{Landa, Landa2, Diego, Velazquez}} 
. {In the shell model, the interaction is typically modeled as the sum of an average-field interaction (monopole field) and an N-body interaction (multipole field) $\hat{H}=\hat{H}_m+\hat{H}_M$. For the sake of simplicity, when some multipole terms are excluded, the interaction strengths must be adjusted to maintain normalization \cite{Dufour}. In its simplest representation, this consists of the sum of an independent-particle field and a two-body interaction. The two-body interaction is usually reduced to the quadrupole and pairing interactions. For our purposes, we omit the pairing interaction, as it has little influence on state mixing, as evidenced by a comparison of this work with that of \cite{Diego}. Thus, we employ a two-body quadrupole interaction (J. Retamosa, 1990) with a free parameter governing its strength.}
It has been shown that, for weak interactions, its energy levels display regular behavior, whereas, as the quadrupole interaction increases, the system undergoes a transition from a regular to a chaotic regime \cite{Diego}. In this way, our interaction is
\begin{equation}
\hat{H}=\hat{H}_{m}-\chi \hat{Q}\cdot\hat{Q},
\end{equation}
whose strength depends on the parameter $\chi$. 

For the case of $^{48}$Ca, within the shell-model framework, the problem is reduced to a $^{40}$Ca core plus eight neutrons in the $fp$ shell. We work with angular momentum and parity {$J^{\pi} = 3^{+}$}, for which 1627 nuclear states are available, and we consider interaction parameters $\chi = 0.01$ and $\chi = 0.25$. These interactions correspond to regular and chaotic behavior, respectively. The diagonalization performed with the Antoine code (Caurier and Nowacki, 1987) provides the energy levels of the available states.

\begin{figure}[H]
\includegraphics[scale=0.5]{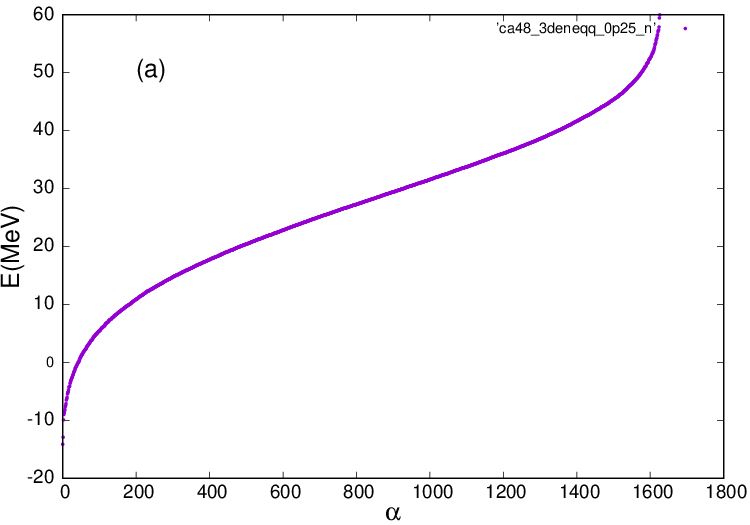} 
\includegraphics[scale=0.5]{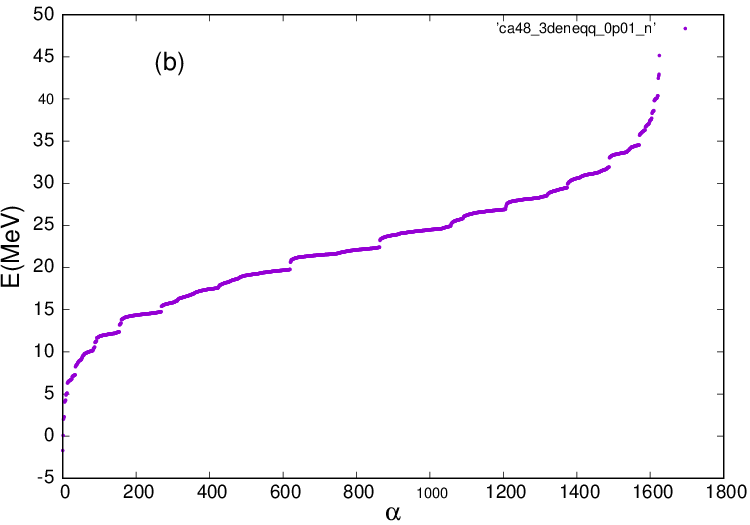}
\includegraphics[scale=0.5]{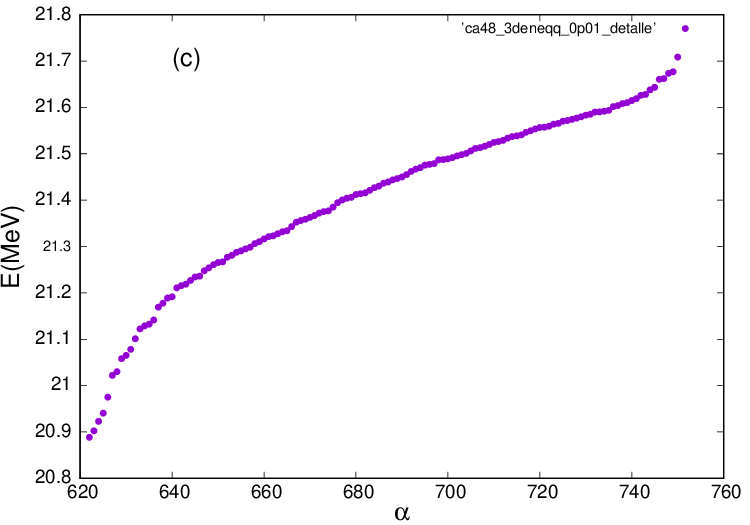} 
\caption{\hl{Energy} 
 levels of \(^{48}\)Ca as a function of the interaction parameter $\chi$. The quadrupole two‑body interaction in (\textbf{a}) $\chi=0.25$ is $25$ times greater than in (\textbf{b}). In (\textbf{c}), a selected sector of the energy spectrum from (\textbf{b}) is shown.}
\label{f1} 
\end{figure}

In Figure \ref{f1}a, we present the sequence of energies obtained from the diagonalization of the Hamiltonian (1) with $\chi=0.01$. The sequence of energies in Figure \ref{f1}a can be approximated by a continuous binomial distribution \cite{Zuker}; however, the second one (Figure \ref{f1}b) shows fluctuations that can be interpreted as a reminiscence of the shell structure.\\
An analysis of the power spectrum of the energy time series shows that the unfolded energy levels for $\chi=0.25$ exhibit noise \(P(f)=1/f\), which is associated with chaotic behavior~\cite{Relano}. For Figure  \ref{f1}b, the energy levels obtained for $\chi = 0.01$ (weak interaction) reveal an internal structural change associated with the formation of correlations between groups of levels, whose power‑spectrum analysis indicates a \(P(f) = 1/f^{2}\) noise, associated with integrable systems \cite{Landa,Relano}. Figure  \ref{f1}c shows a sub‑distribution of the energies for $\chi = 0.01$, corresponding to the subset of energies in Figure  \ref{f1}b with $622 < \alpha < 750$. The power spectrum for this subset of energy levels displays $P(f) = 1/f$ noise. This implies scale invariance of the chaotic behavior.

Thus, the distribution with weak interaction is composed of several sub‑distributions of chaotic energy levels; however, the total power spectrum is not chaotic. In the limit \(\chi = 0.0\), the energy‑level structure is related to the harmonic oscillator with some monopole interactions that weakly break the degeneracy, leaving bunches of energy levels. Each bunch of levels is locally chaotic because there is interaction between all states inside the bunch. When the two‑body interaction increases, states from different bunches interact, leading to a smoother distribution with an overall chaotic behavior characterized by $P(f) = 1/f$.

On the other hand, the chaotic behavior of the system is reflected in the energy-level statistics, which are described by a Wigner distribution. In similar research \cite{Diego}, it has been found that atomic nuclei behave as a chaotic system when the interaction parameter takes values above \(\chi = 0.04\). Under this premise, we calculate the Shannon entropy and study the chaotic behavior in terms of the information content and complexity of the nuclear states.

\section{Shannon Entropy of Nuclear States}

In quantum systems, the problem becomes more difficult because the concept of a trajectory has no meaning in quantum mechanics. However, the dynamics of the system are described by the average behavior (in statistical terms) of each of the elements, and it is here that it makes sense to consider the Shannon entropy for the analysis of a system.  As stated by Zelevinsky,  “The information entropy of individual eigenvectors turns out to be a convenient measure of the degree of complexity of individual wave functions” \highlighting{\cite{Zelevinsky}. } 

In this version of the shell model with the ANTOINE code \cite{Antoine}, the total $J$ angular momentum is projected in the M-direction, which means the M-scheme. In such a case, $J$ total is not a good quantum number that allows us to describe the eigenvectors as $\ket{J;\alpha}=\sum C^{\alpha}_k \ket{J;k}$; however, the projection allows us to argue valid conclusions. For example, the diagonalization in this projected basis $\ket{k}$ gives us the eigenvalues $E_{\alpha}$ with the corresponding eigenvectors $\ket{\alpha}$. It is clear that $E_{\alpha}$ is basis independent; then the amplitudes $C^{\alpha}_k$, depending on the basis, in this case of a projected basis, must have information about the collectivity imposed by the quadrupole--quadrupole interaction. We describe the nucleus \(^{48}\)Ca in terms of a \(^{40}\)Ca core, so that the basis states \(\ket{\alpha}\) of \(^{48}\)Ca are expressed in terms of the basis \(\ket{k}\) of \(^{40}\)Ca as
\begin{equation}
    \ket{\alpha} = \sum_{k} C^{\alpha}_{k} \ket{k}
\end{equation}

\begin{figure}[H]  
\includegraphics[width=1\linewidth]{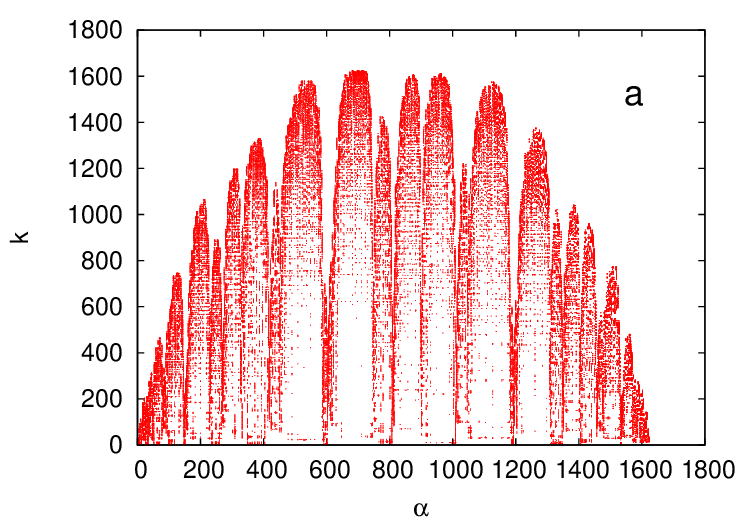}\\
\includegraphics[width=1\linewidth]{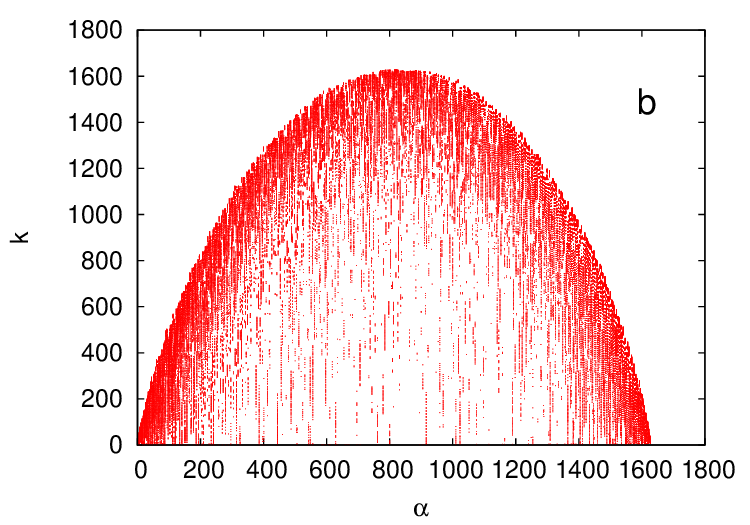}
\caption{{Distribution of k states for which the probabilities $(P_k^{\alpha} = |C_k^{\alpha}|^2>0.05$  for each state vector $|\alpha\rangle$ on the horizontal axis for the \(^{48}\mathrm{Ca}\) nucleus with \(J^{\pi} = 3^{+}\)}. We use two values of the quadrupole–quadrupole interaction: (\textbf{a}) \(\chi = 0.01\) and (\textbf{b}) \(\chi = 0.25\). The probability distributions show gaps for weak interactions.
 }
\label{2} 
\end{figure}

In Figure \ref{2}, we plot each of the 1627 \(k\)  {amplitudes} of the probability \(P_k^{\alpha} = |C_k^{\alpha}|^2\) (on the vertical axis) for each \(|\alpha\rangle\) state (on the horizontal axis), for (a) \(\chi = 0.01\) and (b) \(\chi = 0.25\) \highlighting{\cite{GFernandez}.} 
 To make the patterns clearer, we display only the probabilities satisfying \(P_k^{\alpha} > 0.05\). For a weak quadrupole interaction, the graph shows gaps between components for certain \(\alpha\) states, whereas for a strong interaction these gaps disappear. As shown in Figure \ref{2}a, subsets of \(\alpha\) states exist for weak interactions that do not mix with other subsets. By increasing the quadrupole interaction [Figure \ref{2}b], mixing between subsets of components occurs, smoothing the distribution of \(k\) components. In other words, a strong interaction produces level repulsion, which is reflected in the mixing of a large set of states with different \(\alpha\) values.

The Shannon entropy is defined as 
\begin{equation}
    S^{\alpha} = -\sum_{k} |C^{\alpha}_{k}|^{2} \ln |C^{\alpha}_{k}|^{2}
\end{equation}
To normalize this entropy, we use the maximum value that the information entropy can reach for a Gaussian orthogonal ensemble (GOE) \cite{Zelevinsky}: 
\begin{equation}
    S_{\mathrm{GOE}} = \ln(0.48\,N)
\end{equation}

\begin{figure}[H]  
\includegraphics[width=0.97 \linewidth]{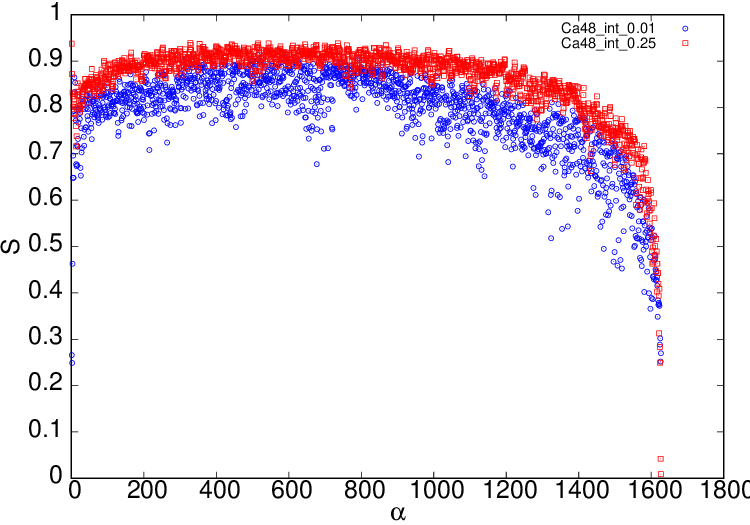}\\
\includegraphics[width=0.97\linewidth]{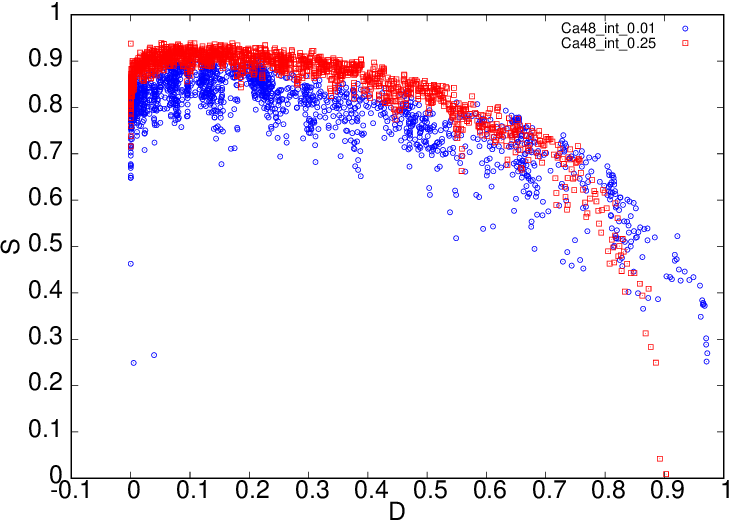}
  \caption{\hl{Shannon entropy} 
 as a function of the interaction parameter for \(\chi = 0.01\) (blue circles) and \(\chi = 0.25\) (red squares). (\textbf{a}) Shannon entropy versus basis state \(\alpha\); (\textbf{b}) Shannon entropy versus disequilibrium \(D\).
 }
    \label{3}
\end{figure}

The results are shown in Figure \ref{3}a, where we can see that the entropy for \(\chi = 0.01\) is on average lower than the entropy for \(\chi = 0.25\). We can also see that for weak interactions, \(\chi = 0.01\), the gaps that appear in Figure \ref{2}a have a counterpart in Figure \ref{3}a in the Shannon entropy for the same interaction, where downward peaks are observed. It is important to mention that for the \(k\) values near the limits \(k = 1\) and \(k = 1627\), the values of \(S\) go to zero because there are few states that are the most probable; these are the most pure states.

Although the graph is intuitive with respect to what we understand by information entropy, it is interesting to adopt another auxiliary quantity to place our system far from or close to order.
We adopt the definition of disequilibrium 
 D' \cite{LopezRuiz}. This is a kind of distance from an equiprobable distribution, defined as 
\begin{equation}
    D'=\sum_{i=1}^N(p_i-\frac{1}{N})^2,
\end{equation}
where $p_i$ is the i-th probability of an N-system, and $1/N$ is the corresponding probability of an equiprobable system. The differences are squared to avoid negative probabilities.

We can plot the entropy versus this distance; however, we prefer to plot the entropy versus the real difference, or distance,
\begin{equation}
D = \sum_{i=1}^{N} \left(p_i - \frac{1}{N}\right)
\end{equation}

By discarding all coefficients with \(|C^{\alpha}_k| < 10^{-8}\), we guarantee that \(\sum_k |C_k^{\alpha}|^2 \approx 1\) for each \(|\alpha\rangle\). In this case, for \(\chi = 0.01\) we obtain 14 cases with \((p_i - \frac{1}{N}) < 0\), in the range \(-10^{-8} \to -10^{-7}\), while for \(\chi = 0.25\) we have 25 such cases in the range \(-10^{-9} \to -10^{-7}\). We expect that these negative differences have very little influence on the results.

When we plot \(S\) vs. \(D\) [Figure \ref{3}b], we identify two limits: at \(D = 0\), corresponding to maximal randomness or maximal entropy, and at \(D = 1\), corresponding to maximal order or minimal entropy. In the same figure, the maximal entropy again occurs for \(\chi = 0.25\), with \(S > 0.9\), but this case reaches an order limit around \(D \approx 0.9\); in contrast, the entropy for \(\chi = 0.01\) is on average lower than for \(\chi = 0.25\), but it can attain states of almost complete order, with \(D\) close to 1.

\section{Complexity as a Tool to Characterize Nuclear Behavior}

Complexity can be interpreted as a measure of the balance between organization and disorganization. It can also be understood in terms of the amount of information needed to describe a system, considering two extreme cases: a crystal and an ideal gas, which require minimal and maximal information, respectively, to be described \cite{LopezRuiz}. Intuitively, the complexity tends to zero when the system is near the two limits \(D = 0\) and \(D = 1\). The maximum of the complexity is reached at some point between these two limits, defining the emerging dynamics.

We define the complexity as 
\begin{equation}
    C = S \cdot D
\end{equation}
with \(S\) normalized by Equation (4). Another common definition is \(C = 4S(1-S)\) \cite{Pineda}, where \(D = 1 - S\) is the disequilibrium. 

\begin{figure}[H]
  \subfloat[\centering]{\includegraphics[scale=0.75]{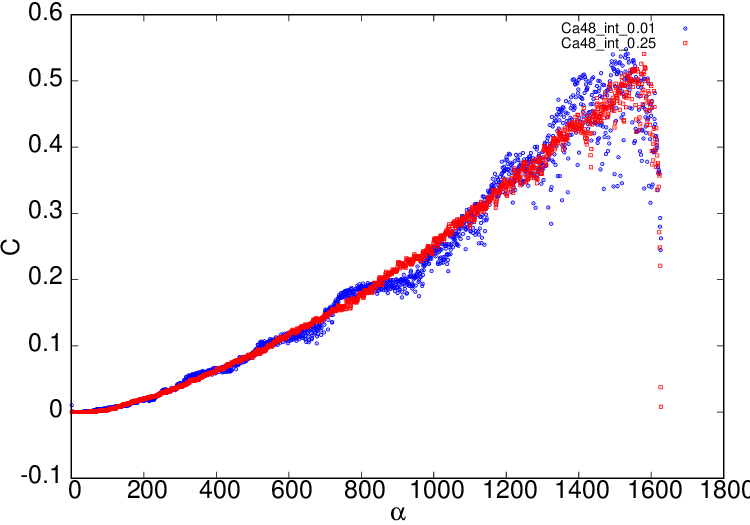}}\\
  \subfloat[\centering]{\includegraphics[scale=0.75]{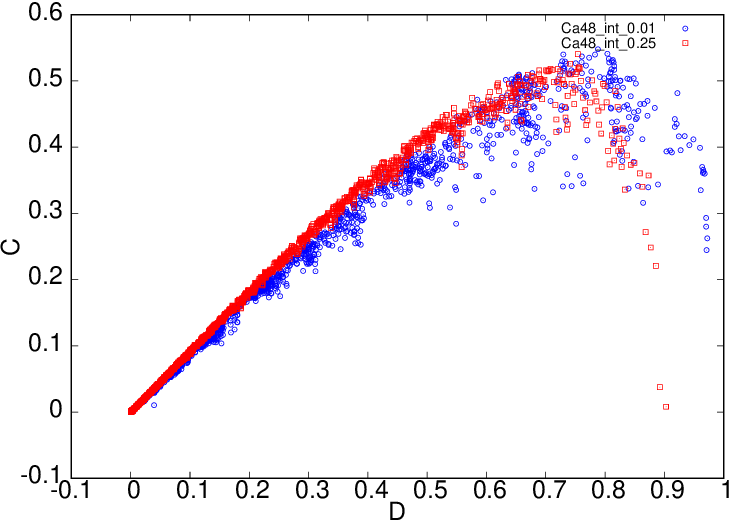}}
\caption{\hl{Complexity} 
 as a function of the interaction parameter for \(\chi = 0.01\) (blue circles) and \(\chi = 0.25\) (red squares). (\textbf{a}) Complexity versus basis state \(\alpha\); (\textbf{b}) Complexity versus disequilibrium \(D\).}
\label{f4} 
\end{figure}

Figure \ref{f4}a shows the complexity \(C\) as a function of \(\alpha\). As with the Shannon entropy, the complexity vanishes at the extremes, where the \(\alpha\) states are almost pure, and increases in the intermediate region where configuration mixing occurs. However, the maximum complexity is not located at the center, so it is useful to examine complexity as a function of disequilibrium. The complexity pattern also shows ripples, with local maxima and minima. These arise from the internal structure of energies and eigenvectors seen in Figures \ref{f1}c and \ref{2}a, since the weak interaction is not strong enough to mix configurations between the different subsets of states.

\newpage
The maxima of both complexity curves are near \(\alpha = 1600\). It is risky to say that the states near the limit are the most complex, since the denationalization performed with ANTOINE in the M scheme does not use the total angular momentum as a good quantum number. However, independent of the specific \(\alpha\) state, with the help of the disequilibrium, we can estimate where the maximum of the complexity is located.

In Figure \ref{f4}b, the complexity depends on the disequilibrium. The structure of both complexity curves for \(\chi = 0.01\) and \(\chi = 0.25\) is similar to that in Figure \ref{f4}a; however, the maximum complexity for \(\chi = 0.25\) appears near \(D = 0.75\), while for \(\chi = 0.01\) it appears near \(D = 0.8\). For the complexity, there are several important values of \(D\): the two values inherent to its definition, \(D = 0\) and \(D = 1\), which correspond to random and ordered phases, or physically to ideal‑gas and crystalline phases, respectively. However, the position of the maximum complexity can be basis‑dependent. In our case, we will emphasize in the conclusions the differences between the complexity limits of the chaotic phase for \(\chi = 0.01\) and \(\chi = 0.25\).

\section{Conclusions}

We studied the entropy and complexity of the \(^{48}\)Ca nucleus for the set of states with \(J = 3\) and positive parity, in the ANTOINE M scheme, and compared Shannon entropy and complexity for quadrupole two‑body interactions with two different strengths, \(\chi = 0.01\) and \(\chi = 0.25\). For the weak interaction, the energies and eigenvectors show clusters or groups of states, while increasing the interaction by a factor of twenty-five mixes most configurations and smooths the spectrum towards a binomial‑like distribution. This increase in interaction produces a transition from \(1/f^{2}\) noise to chaotic \(1/f\) noise. Comparing the entropy for the two interactions, we see that the stronger interaction gives a more uniform and higher entropy, confirming that the system with \(1/f^{2}\) noise is more organized than the system with \(1/f\) noise. Since complexity depends directly on entropy, the qualitative behavior is similar: increasing the quadrupole interaction, which largely controls the nuclear shape, enhances configuration mixing and leads to greater complexity. Interestingly, we can also observe local complexity maxima associated with the weakly interacting clusters or groups of states in the spectrum, suggesting that, like entropy, complexity can be discussed in terms of contributions from such groups.

It is also important to note that the dependence of complexity on disequilibrium explains the slight separation between the global complexity maxima for the two interactions. The complexity maximum for \(\chi = 0.01\) is closer to the ordered limit \(D = 1\), while for \(\chi = 0.25\) it occurs at a smaller \(D\), before approaching \(D = 0.9\). 

For systems such as the atomic nucleus, we know that a maximum of complexity is not found at the extreme limits, but at an intermediate point. The position of this maximum may be close to \(D = 3/4\), as suggested by Figure \ref{f4}b. What we can firmly state from our results is that complexity is sensitive to the structure of the interactions. Since the energies are basis‑independent, the presence of level groups with chaotic structure [Figure \ref{f1}b] is reflected in local complexity maxima. This implies that, for a larger quadrupole interaction that mixes configurations between different groups of states, a single global complexity maximum emerges. Therefore, the \(1/f\) noise associated with quantum chaos corresponds to a single complexity maximum.







\vspace{6pt} 
\authorcontributions{Conceptualization, L.L. and V.V.; methodology, L.L.; validation, V.V., C.V.; formal analysis, D.L.;investigation, L.L. and D.L.; writing---original draft preparation, L.L.; writing---review and editing, V.V. and C.V.;supervision, C.V. and V.V.;  } 

\funding{This research received no external funding} 

\dataavailability{Data can be requested from vicvela@ciencias.unam.mx } 

\acknowledgments{
 LLH gratefully acknowledge a CONAHCYT graduate scholarship.}

\conflictsofinterest{The authors declare no conflicts of interest.} 
\end{document}